\documentclass{ws-procs975x65}

\begin{document}
\title{ACCESSING THE ACCELERATION OF THE UNIVERSE WITH SUNYAEV-ZELDOVICH AND X-RAY DATA FROM GALAXY CLUSTERS }
\author{RODRIGO F. L. HOLANDA, JOS\'E A. S. LIMA$^*$, JO\~{A}O V. CUNHA}
\address{Departamento de Astronomia, Universidade de S\~{a}o
Paulo \\ Rua do Mat\~ao, 1226 - 05508-900, S\~ao Paulo, SP, Brazil, \\$*$E-mail:limajas@astro.iag.usp.br }
\begin{abstract}
By using exclusively the Sunyaev-Zel'dovich effect and X-ray
surface brightness data from 25 galaxy clusters in the redshift
range $0.023\leq z \leq 0.784$  we access cosmic acceleration employing a kinematic
description. Such result is fully independent on the validity of
any metric gravity theory, the possible matter-energy contents
filling the Universe, as well as on the SNe Ia Hubble diagram. 
\end{abstract}
\bodymatter
\section{Introduction}
Currently, SNe type Ia  provides a unique direct access to the late
time accelerating stage  of the Universe \cite{RIESS98}. Naturally, this a rather
uncomfortable situation from the observational and theoretical
viewpoints. A promising distance estimator fully independent of SNe type Ia and other
calibrators of the cosmic distance ladder is the angular diameter
distance ($D_{A} (z)$) from a given set of distant objects\cite{Kellerman93}. It has
also been recognized that the combination of SZE\cite{sunzel70} and
X-ray surface brightness measurements  may provide useful angular
diameters from  galaxy clusters
\cite{caval77,filippis06,CML07}.

De Filippis et al. (2005)
\cite{filippis06} reanalyzed and derived, using an isothermal
elliptical 2-Dimensional $\beta$-model to describe the clusters,
${\cal{D}}_A$ measurements for 25 clusters from two previous
compilations \cite{Reese02,Mas01}, where was used a spherical isothermal $\beta$ model to
describe the clusters geometry. More recently, it was shown \cite{HLR,HLR2} that De Filippis {\it{et al.}} sample is in good agreement with
the distance duality (DD) relation between luminosity distance
(${\cal{D}}_L$) and angular diameter distance (${\cal{D}}_A$),
${\cal{D}}_L(1+z)^{-2}/{\cal{D}}_A=1$, in the context of a $\Lambda$CDM model
(WMAP7) \cite{komatsu}.  This sample is also consistent (2$\sigma$ c.l.)
with no violation of the DD relation in a model-independent cosmological
test involving $D_A$ from galaxy clusters and $D_L$ from the
supernovae Ia data  provided by the Constitution compilation \cite{hicken}. 

In this work, we investigate the potentialities
of SZE/X-ray technique by employing a purely kinematic description of the
universal expansion based on angular diameter distances of clusters
for two different expansions of the deceleration parameter\cite{CL08} : $q(z) = q_{0} +  q_{1}z$ and $q(z) = q_{0} +  q_{1}z/(1+z)$.  As we
shall see, by using the De Filippis {\it et al.}\cite{filippis06} sample
we find that a kinematic analysis based uniquely on cluster data
suggests that the Universe  is accelerating today, such as, $q_{0} <0$ with 83\% of probability.
 
\section{ Kinematic Approach and Constraints} 

Let us now assume
that the Universe is spatially flat as motivated by inflation and
WMAP measurements \cite{komatsu}. In this case, the angular diameter
distance in the FRW  metric is defined by (in our units $c=1$),
\begin{eqnarray}\label{eq:dLq}
D_A &=& (1+z)^{-1}H^{-1}_{0}\int_0^z {du\over H(u)} = \frac{(1+z)^{-1}}{H_0} \nonumber \\
&& \,\, \int_0^z\, \exp{\left[-\int_0^u\, [1+q(u)]d\ln
(1+u)\right]}\, du,
\end{eqnarray}
where  $H(z)=\dot a/a$ is the Hubble parameter, and, $q(z)$, the
deceleration parameter, is defined by

\begin{eqnarray}\label{qz}
q(z)\equiv -\frac{a\ddot a}{\dot a^2} = \frac{d H^{-1}(z)}{ dt} -1.
\end{eqnarray}

Let us now consider the 25 measurements of angular diameter
distances  from galaxy clusters as obtained through SZE/X-ray method
by De Filippis and coworkers \cite{filippis06}. In our analysis we use a
maximum likelihood determined by a $\chi^{2}$ statistics

\begin{equation}
\chi^2(z|\mathbf{p}) = \sum_i { ({\cal{D}}_A(z_i; \mathbf{p})-
{\cal{D}}_{Ao,i})^2 \over \sigma_{{\cal{D}}_{Ao,i}}^2 +
\sigma_{sys.}^{2}}, \label{chi2}
\end{equation}
where ${\cal{D}}_{Ao,i}$ is the observational angular diameter
distance, $\sigma_{{\cal{D}}_{Ao,i}}$ is the uncertainty in the
individual distance, $\sigma_{sys}$ is the contribution of the
systematic errors and the
complete set of parameters is given by $\mathbf{p} \equiv (H_{0},
q_{0},q_{1})$. We have marginalized on the Hubble Distance ($H_{0}^{-1}$) with a gaussian prior on $H_{0}$ centered in our best fit value $H_{0}=77 \pm 4$ km/s/Mpc.
\section{Results and Discussion}

In Fig. 1(a) we show  the contour in the plane $q_{0}-q_{1}$ for linear parametrization.
The confidence region (1$\sigma$) are $-3.2 \leq q_0 \leq 0.7$ and $29 \leq q_1
\leq -9$. In Fig. 1(b) we show the likelihood  for $q_{0}$ (solid line-only statistical errors and dashed line-statistical + systematic errors). We have marginalized on the Hubble distance with a gaussian prior on $H_{0}$ centered in its best fit $H_{0}=77 \pm 4$  km/s/Mpc and on all values of $q_{1}$. We obtain that $q_{0}<0$ with 84\% of probability, $q_{0}=-1.29^{+ 1.31}_{-1.5}$ (1$\sigma$ - only statistical errors) and 74\% of probability, $q_{0}=-0.8^{+ 1.6}_{-1.7}$ (1$\sigma$ - statistical + systematic errors). In Figs.
1(c) and 1(d) we display the corresponding plots for the non linear
parameterization. The confidence region (1$\sigma$) is now defined
by: $-3.5 \leq q_0 \leq 0.9$ and $-17 \leq q_1 \leq 27$. In Fig. 1(d) we show the likelihood distribution function  for $q_{0}$. We obtain that $q_{0}<0$ with 82\% of probability, $q_{0}=-1.4^{+1.41}_{-1.6}$ (1$\sigma$ - only statistical errors) and  with 72\% of probability, $q_{0}=-1^{+1.7}_{-1.6}$ (1$\sigma$ -  statistical + systematic  errors).
\begin{figure*}[h!]
\centerline{\psfig{figure=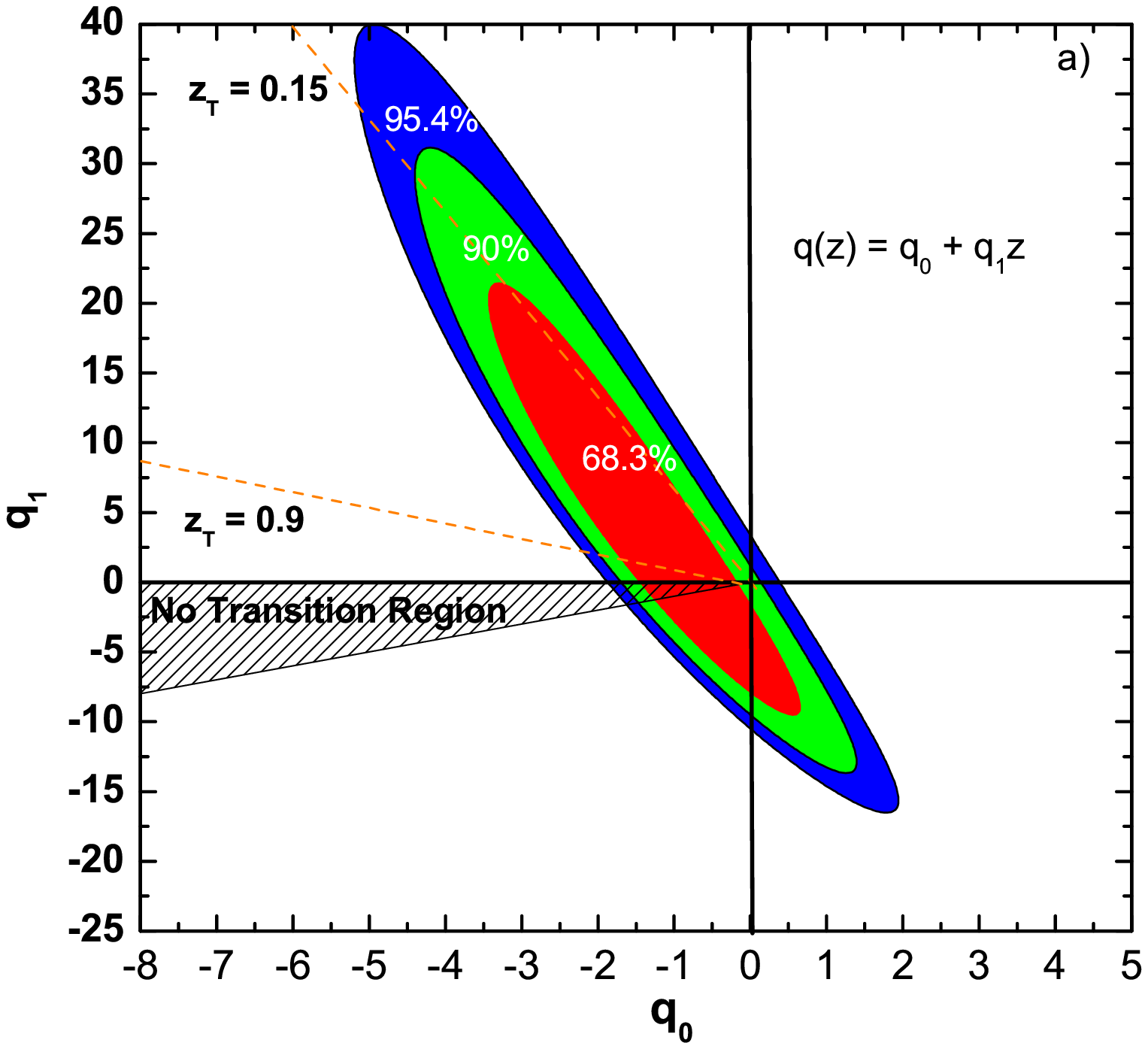,width=1.5truein,height=1.5truein}
\psfig{figure=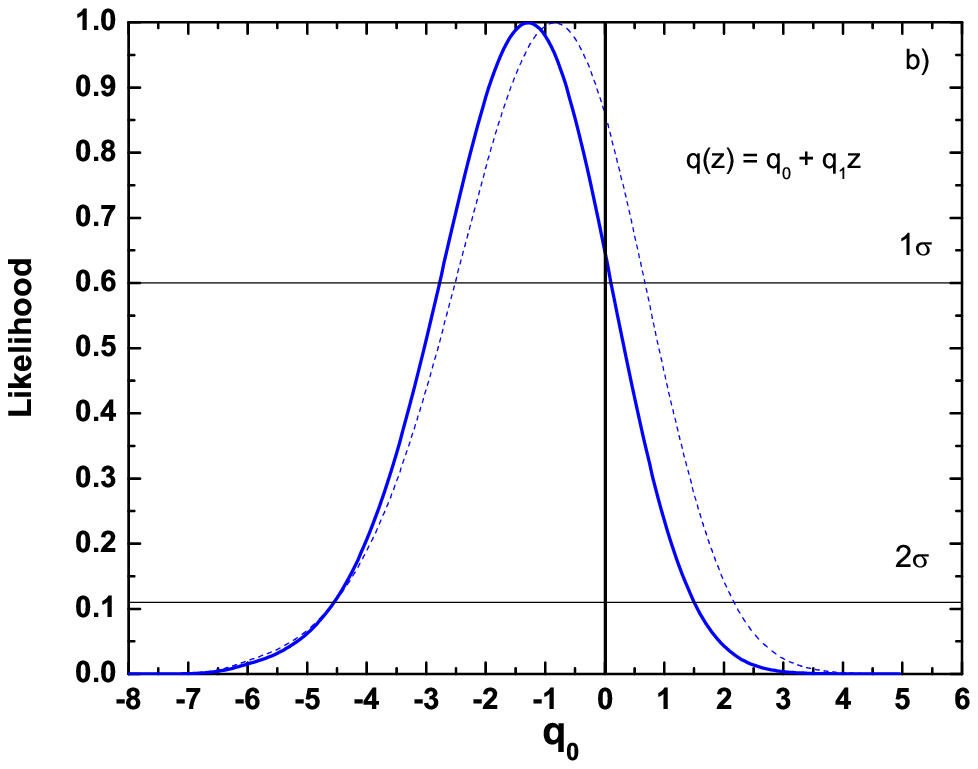,width=1.5truein,height=1.5truein}
\psfig{figure=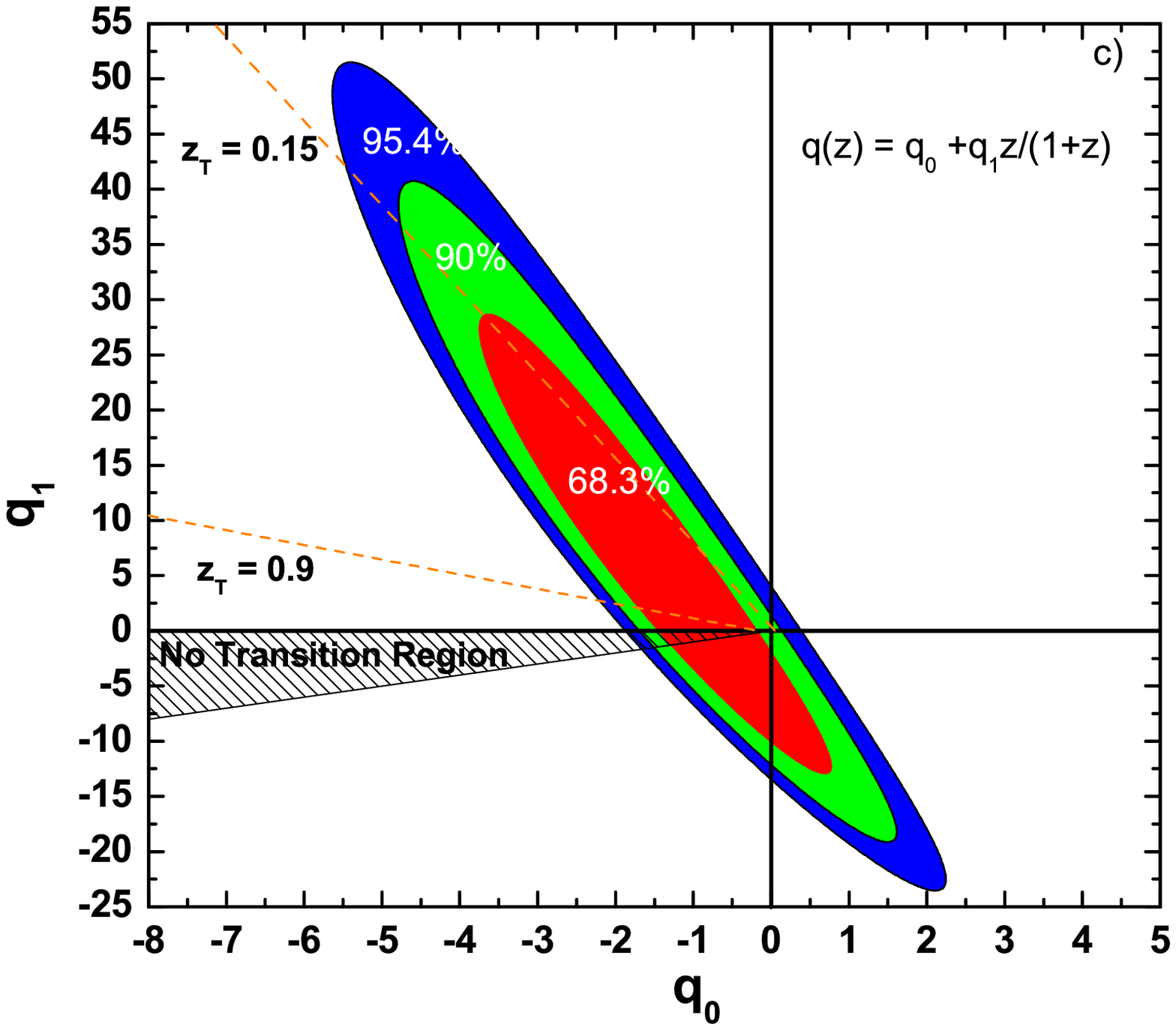,width=1.5truein,height=1.5truein}
\psfig{figure=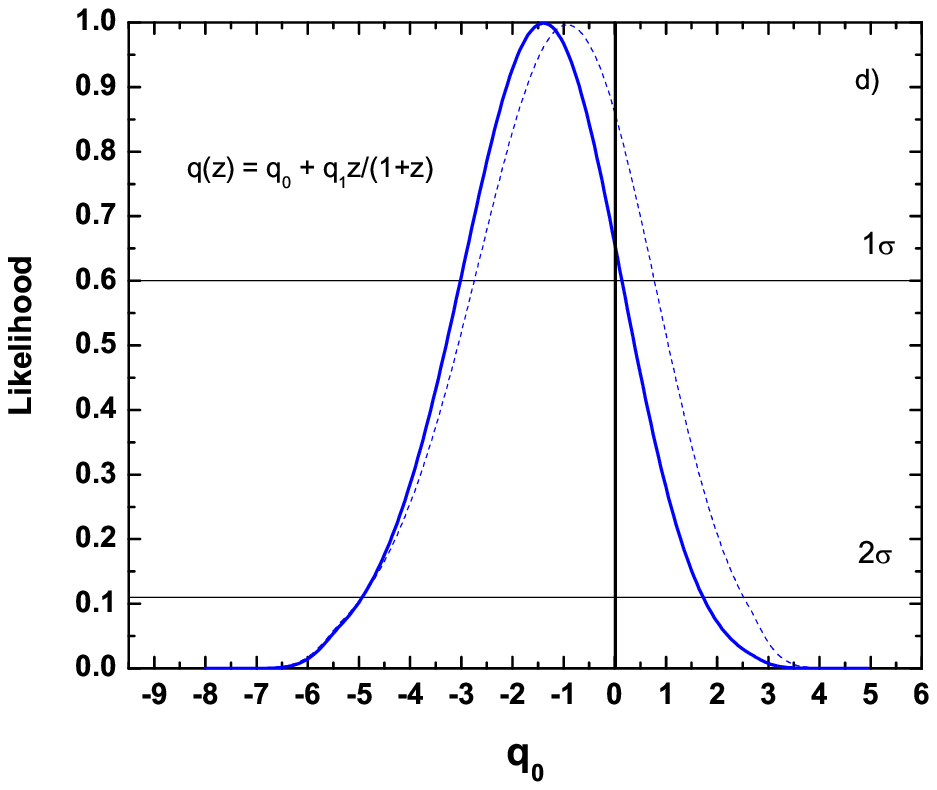,width=1.5truein,height=1.5truein}
 \hskip
0.1in} \vspace{0.5cm}
\caption{ {\bf{a)}} Contours in the $q_o - q_1$ plane for 25 galaxy
clusters data \cite{filippis06} considering $q(z)=q_0 +q_1z$. {\bf{b)}} Likelihood function for the
$q_{0}$, marginalizing on all values of $q_{1}$. {\bf{c)}} Contours in the $q_o - q_1$ plane  considering $q(z)=q_0 +q_1z/(1+z)$. {\bf{d)}} Likelihood function for the
$q_{0}$, marginalizing on all values of $q_{1}$ for non-linear parametrization.}\label{fig1}
\end{figure*}

We have shown that the combination of
Sunyaev-Zeldovich/X-ray data from galaxy clusters is an interesting
technique for accessing the present accelerating stage of the
Universe.  This result follows from a consistent kinematic
approach based on the angular diameter distance of galaxy clusters
obtained from SZE/X-ray measurements. By using two different
parameterizations, it was found that $q_{0}<0$  with at least $83\%$ probability (only statistical errors in galaxy clusters data) and $72\%$ probability (statistical + systematic errors in galaxy clusters data).

\end{document}